\documentclass[aps,twocolumn]{revtex4}

\usepackage{graphicx,amssymb}

\begin{document}

\title{Staggered flux vortices and the superconducting transition in the layered cuprates}
\author{Carsten Honerkamp and Patrick A. Lee}
\affiliation{Department of Physics, Massachusetts Institute of Technology, Cambridge MA 02139, USA }  
\date{September 19, 2003}

\begin{abstract} 
We propose an effective model for the superconducting transition in the high-$T_c$ cuprates motivated by the SU(2) gauge theory approach. 
In addition to variations of the superconducting phase we allow for local admixture of staggered flux order. 
This leads to an unbinding transition of vortices with staggered flux core that are energetically preferable to conventional vortices. 
Based on parameter estimates for the two-dimensional $t$-$J$-model we argue that the staggered flux vortices provide a way to understand a phase with a moderate density of mobile vortices over a large temperature range above $T_c$ that yet exhibits otherwise normal transport properties. This picture is consistent with the large Nernst signal observed in this region.
\end{abstract}
\maketitle

The nature of the pseudogap phase of the (hole-)\-underdoped high-$T_c$ cuprates is one of the central questions of correlated electron physics. 
A number of scenarios and descriptions are successful in capturing certain aspects of the problem. Yet many theories face substantial difficulties when it comes to combining the large number of experimentally established anomalies of the underdoped state. A new challenge in this context has been set out recently by the Nernst effect measurements in underdoped samples\cite{wang}. In these experiments a thermal gradient is applied in the copper-oxide planes. In the presence of a small out-of-plane magnetic field, a voltage drop in the direction perpendicular to magnetic field and thermal gradient is observed. This voltage is interpreted as the phase slip signal arising from vortices moving from hot to cold. Thus the Nernst effect experiment reveals the existence of substantial superconducting (SC) short range correlations over a sizable region, starting significantly below the onset temperature $T^*$ of the pseudogap, but extending up to temperatures high above the low $T_c$s of underdoped samples. 

At first sight the observation of vortices  above $T_c$ fits well into a scenario\cite{emkiv} where SC phase fluctuations destroy the long range coherence, and short range pairing correlations survive up to much higher temperatures. In view of the small superfluid weight of underdoped systems it is conceivable that the SC transition is driven by a vortex unbinding similar to the XY-transition, the critical temperature disappearing like the doping $x$ for $x \to 0$. 
However a simple phase fluctuation scenario faces the following problem. The creation of a vortex comes at a price, as the SC order parameter goes to zero in the vortex core and condensation energy is lost. In disordered films of conventional superconductors, where a Berezinskii-Kosterlitz-Thouless(BKT) transition can be observed, the mean-field critical temperature and the vortex unbinding temperature are similar, and due to the mean free path $\ell$ entering the effective coherence length $\xi =\sqrt{\xi_0 \ell}$ (with $\xi_0 \sim v_F/\Delta$) 
the vortex core energy $\sim \xi^2 \Delta^2/\varepsilon_F$ becomes small near $T_c$, as $\Delta$ decreases faster for $T\to T_c$ than $\xi$ increases. Thus the loss of condensation energy is limited and the vortices can be cheap. 
In contrast with that the high-$T_c$ cuprates are in the clean limit and for underdoped samples the gap magnitude remains large up to temperatures far above the SC transition. 
Hence the vortex core energy for bringing the gap magnitude down to zero inside the vortex would normally be expected to be huge ($\varepsilon_F$ in BCS theory and of order $J$ in our case), and only exponentially few of these expensive vortices could be created above the small $T_c$ in underdoped samples. Then we would expect that the transport properties of the pseudogap state resemble those of a flux-flow (FF) phase. Transport resembling the normal state only occurs at high temperatures of order of the core energy when the vortices proliferate and overlap.  
However this picture is inconsistent with experiments. These show that above a limited fluctuation regime close to $T_c$ the in-plane transport  looks rather normal and signs of FF conductivity do not extend far above $T_c$ -- contrary to the Nernst signal. Apparently the conductivity $\sigma_{n}$ due to a significant number of quasiparticles dominates over the FF conductivity $\sigma_{\mathrm{FF}}$ in the total conductivity $\sigma=\sigma_n+\sigma_{\mathrm{FF}} $. Thus one has to explain two things: where the normal excitations come from and why the FF contribution is small. 
$\sigma_{\mathrm{FF}}$ can be estimated to be $\propto \eta / n_V $, where  $n_V$ is the density of vortices either forced in by a magnetic field  in the mixed state or generated thermally above the BKT transition. $\eta$ is the friction coefficient for the vortex motion. The FF conductivity is small if $\eta$ is small and $n_V$ is not. 
In the following we show that our model produces a moderately large  $n_V$ even for underdoped samples with low $T_c$. We will also present an approximate calculation 
that yields a finite density of normal excitations. 
We will not attempt to calculate $\eta$ in this work. An effect that may reduce $\eta$ is the experimentally observed\cite{vortexcore} small low-energy density of states in the vortex cores. 
This translates into small dissipation due to vortex motion and thus small $\sigma_{\mathrm{FF}}$. Note that $\eta$ cannot become arbitrarily small if we want $\sigma_n$ to dominate the conductivity, as the the quasiparticles responsible for  $\sigma_n$ will cause some dissipation.
Thus a small FF contribution requires a moderately large $n_V$ at temperatures above the limited fluctuation regime near $T_c$.     

An extreme way to obtain normal transport properties just above $T_c$ is to make the vortices very dense, such that they overlap just above the transition. Then it is hard to understand why the vortex Nernst signal persists to temperatures so high above $T_c$. 
A theory considering purely Gaussian SC fluctuations\cite{ussishkin} gave good agreement for the Nernst effect in overdoped and optimally doped samples.  However the description of underdoped samples becomes problematic and an additional suppression of $T_c$ had to be invoked.
 
Thus neither very few and expensive nor too many and too cheap vortices seem to match the experimental picture. Basically what is needed is a theory which produces a core energy of order $T_c$ rather than $J$. 
In other words we need to decrease the core energy by placing in the vortex core another non-superconducting state that is nearby in energy. 
There are several proposals in the literature\cite{arovas} for such a cheap vortex core, mainly emphasizing the vicinity of the $d$-wave superconductor to other ordered states. Here
 we study the possibility of a staggered flux (SF) state inside the vortex\cite{su2v,dhlee}. This scenario has the advantage that it emerges naturally from the SU(2) invariance, i.e. the Mott insulating nature of the undoped state. Note that the vicinity to the Mott state is also responsible for the small superfluid stiffness $\rho_s \sim x$.

The idea that vortices in the underdoped system have SF cores ties in with a more general picture of the pseudogap state. This is derived from the SU(2) gauge theory for the $t$-$J$ model and views the pseudogap regime as a thermally disordered state, where the system fluctuates between various types of short range order corresponding to mean-field states that would all become identical at zero doping. The two most prominent correlations are $d$-wave superconductivity - determining the ground state as soon as the other fluctuations freeze out at low $T$ -- and SF correlations. The latter represent - in addition to conventional phase fluctuations of the SC order parameter - the lowest lying fluctuations around the $d$-wave SC state with the largest spectral weight\cite{collmodes}. We have argued\cite{honlee} that the scattering of quasiparticles with these SF fluctuations may be related to the partial loss of the quasiparticle peaks in the pseudogap state. 

Some of the ideas presented here carry over to other types of cheap vortices. Indeed there are experimental indications\cite{kaku} for antiferromagnetic ordering at low temperatures in the vortex cores of optimally doped Tl-compounds, and it is quite likely that the vortex core may contain both  SF and AF  correlations\cite{ogata}. The focus on SC and SF correlations is an attempt to concentrate on the main tendencies suggested from the SU(2) approach. The SF state has the advantage that it has a gap structure identical to the $d$-wave superconductor and naturally explains the gap in the vortex core. 
Other types of correlations, such as antiferromagnetic tendencies, may be viewed as additional instabilities, which naturally coexist with the SF order for small doping.

Recently Ivanov and Lee\cite{dima} calculated the energy differences between SF and $d$-wave SC states using the Gutzwiller projection technique. 
Together with the computed superfluid stiffness this provides an estimate for the energy of a SF vortex. 
Note that in Ref. \cite{dima} the vortex core turns out to very small, but on the other hand the core state, taken to be a pure SF state, is likely to be somewhat too high in energy, and a better core state will increase the size of the core again.  
Keeping in mind these uncertainties and for the lack of better parameters we will use the numbers of Ref. \cite{dima} as input for a generalized XY-model. 

Let us begin with the SU(2) mean-field (MF) theory\cite{collmodes}. Here we are interested in low temperatures. Hence we assume that the bosons carrying the electronic charge are condensed. 
The  Hamiltonian for the fermionic spin degrees of freedom $f_{i\uparrow}$, $f_{i \downarrow}$ on the lattice sites $i$ 
is given by 
\begin{equation} H_f= \frac{J}{2} \sum_{\langle ij \rangle} 
\left( \begin{array}{c} f_{i\uparrow} \\ f^{\dagger}_{i\downarrow} \end{array} \right)^{\dagger} \hspace{-1mm}\left( \begin{array}{cc} -\chi + W_{ij} & \Delta_{ij} \\ \Delta^*_{ij}&  \chi + W_{ij} \end{array} \right) \hspace{-1mm}
\left( \begin{array}{c} f_{j\uparrow} \\ f^{\dagger}_{j\downarrow} \end{array} \right) \label{fermions}
 \, . \end{equation}
 $\chi$ contains the hopping contributions and the constraint field $a_{0,3}$, which is a function of the doping. 
Next we allow for local admixtures of SF amplitude in exchange for SC pairing amplitude, described by an angle $\theta_i$\cite{collmodes}, 
and fluctuations of the SC phase, $\alpha_i$.
Then the SF amplitude on the bond $ij$ is given by 
$W_{ij} = i \Delta_0 (-1)^{i_x+j_y} \cos \frac{\theta_i+\theta_j}{2}$ and the pairing amplitude is $\Delta_{ij}=  (-1)^{i_y+j_y} \Delta_0 \sin \frac{\theta_i+\theta_j}{2} \exp \left[ i (\alpha_i +\alpha_j)/2 \right]$. 
The pure superconductor has $\theta=\pi/2$, while the two degenerate SF states have $\theta=0$ and $\theta=\pi$. 

Now consider an effective Hamiltonian for $\alpha_i$ and $\theta_i$,
\begin{eqnarray}    H &=&  \sum_{\langle ij\rangle} \rho_s(x,\theta_i,\theta_j) \, \cos (\alpha_i-\alpha_j) + \sum_{i \notin \mbox{\tiny VP}} m_{\theta} \cos^2 \theta_i \nonumber \\ && +  K \sum_{\langle ij \rangle} \left( \theta_i - \theta_j \right)^2 + \sum_{\mbox{\tiny VP}} H_V \, . \label{effham} \end{eqnarray}
The first term is the phase stiffness of the SC phase that depends on doping $x$ and the local $\theta$. The second term is a mass term for $\theta$ that takes into account the energy difference between SF and SC state outside the vortex plaquettes (VP).
The third term is a gradient term for the $\theta$-variation. 
The last term is the vortex core energy, $H_V = \sum_{i \in \mbox{\tiny VP}} (  m_n \sin^6 \bar{\theta} + m_\theta)$. 
The sum is over the four sites on each of the VP. Here $\Delta_{ij}$ is assumed to vanish and according to Eq. \ref{fermions}, $\theta$ now describes an interpolation between the SF state ($\theta=0$,$\pi$) and the zero flux state ($\theta=\pi/2$) which is a Fermi liquid. As discussed earlier, the Fermi liquid core is expected to be costly and the energy costs of the SF state and the Fermi liquid state are $m_\theta$ and $m_\theta+m_n$, respectively.
The specific $\theta$-dependence chosen for $H_V$ comes from the MF theory for uniform $\theta$ and $\bar{\theta}$ is the average $\theta_i$ over the four sites of the VP.
Requiring that the vortex core area $4a^2$ equals the vortex size $\pi \xi^2$ of a pure SF vortex in the microscopic calculation in Ref. \cite{dima}, we obtain that the lattice constant of Eq. \ref{effham} is slightly larger than the one of the underlying $t$-$J$ model with a scale factor of $\sim 1.2$. 
We assume that the variation of the vortex core size can be neglected in the small doping and low temperature region we are interested in. 

In principle, all parameters for (\ref{effham}) can be obtained from the MF theory by integrating out the fermions as in Ref.\cite{collmodes}, 
but where possible we use parameters obtained by  
the Gutzwiller variational treatment of Ivanov and Lee\cite{dima}. 
With $t/J=3$, this gives $\rho_s(x,\theta=\pi/2)\approx 0.75xJ$, $m_\theta \approx 0.33xJ$ and $m_n \approx (0.25-x)J/2$. The $\theta$-dependence of $\rho_s$ is obtained numerically from MF theory which shows a rapid linear increase from zero when $\theta$ deviates from $0$ or $\pi$.
The coefficient $K$ for the gradient of $\theta$ is difficult to extract from the  collective mode spectrum\cite{collmodes}, as the $\theta$-mode energy is not simply quadratic around $(\pi, \pi)$. Nevertheless the $q$-dependence of the $\theta$-variations is quite weak and less than that of the SC phase $\alpha$ and goes to zero for $x \to 0$. Thus we estimate $K \sim \rho_s/2$. Other choices give similar results.

\begin{figure}
\includegraphics[width=.49\textwidth]{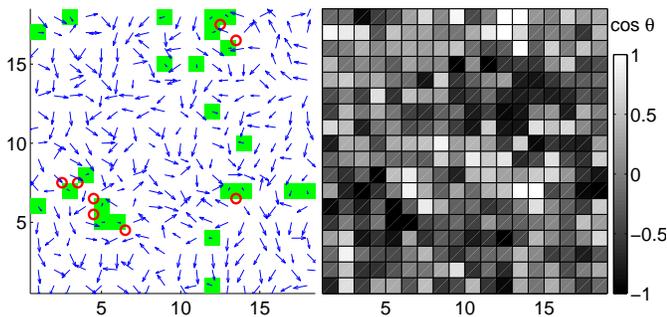}

\caption{Snapshot of the simulation for $x=0.06$ and $T=0.09 J \approx 2 T_c^e$.  Left: Vortices (circles) and phase angles (arrows). The arrow length denotes the local SC amplitude. 
The shaded squares indicate sites with large SF admixture, $|\cos \theta_i|>0.9$. 
The average value of $|\cos \theta |$ in the vortex cores is 0.7 (1 for the pure SF state). Right: SF fluctuations $\cos \theta_i$ of the same sample.}
\label{sam}
\end{figure} 
In the scaling theory for the BKT transition\cite{chailub} the vortex core energy  is assumed to be large compared to the temperature such that the fugacity $y = \exp ( -E_c/T )$ can be used as small parameter. In the limit $y \to 0$ the transition occurs at $T_c =\pi \rho_s/2$. 
A nonzero fugacity $y>0$ leads to a reduction of $T_c$ from this upper bound 
by roughly $\pi^2 \rho_s y$.  With the parameters above the core energy of a 
single ideal SF vortex is  $E_c= 0.75 \pi x J$ or twice the maximal $T_c^{max}=\pi\rho_s /2$. This leads to a reduction of the critical temperature down to $T_c \approx 1.06 \rho_s$. 

The model described by Eq. \ref{effham} can be simulated with Monte Carlo methods. To estimate $T_c$ we calculate the helicity modulus $\Upsilon$ 
which\cite{teitel} measures the rigidity with respect to a phase 
gradient in the system.  $\Upsilon$ vanishes above the SC transition: in BKT theory it jumps to zero at $T_c$, the height of the jump being $2T_c/\pi$, independent of the core energy\cite{nelson}. 
We also measure the average $\theta$-variation inside and outside the vortex cores and how the number of vortices depends on doping and temperature. 

\begin{figure}
\includegraphics[width=.49\textwidth]{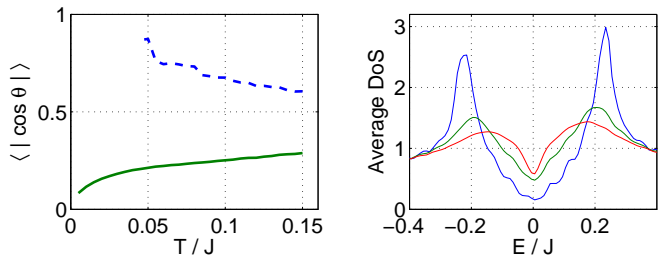}
\caption{Left: SF admixture inside (dashed line) and outside (solid line) the vortices vs. temperature $T$ for $x=0.06$. 
Right: Density of states for $x=0.06$ and $T=0.01$, $0.05$, and $0.13J$, averaged over 25 samples (sizes 40$\times$40 up to 48$\times$48).}
\label{vcont}
\end{figure}

A snapshot from the Monte Carlo is shown in Fig. \ref{sam} for a sample above $T_c$. 
We observe two relatively well isolated and other less separated vortices. 
The SC phase $\alpha_i$ is disordered but  exhibits remnants of short range order. 
For $T=0.09J$ and $x=0.06$, $\theta$ varies rapidly in space due to its light mass $m_\theta <T$ and small gradient terms. Hence there is a significant amount of SF admixture reducing the SC pairing amplitude locally even outside the vortex cores (see shaded areas in Fig. \ref{sam}). Outside the vortices, the average SF amplitude is $\langle |\cos \theta| \rangle \approx 0.3$ (compared to 1 for the pure SF state). Inside the vortex cores it is strongly enhanced, but not maximal ($\langle |\cos \theta| \rangle \approx 0.7$ for the sample shown, see also left plot in Fig. \ref{vcont}).  

\begin{figure}
\includegraphics[width=.49\textwidth]{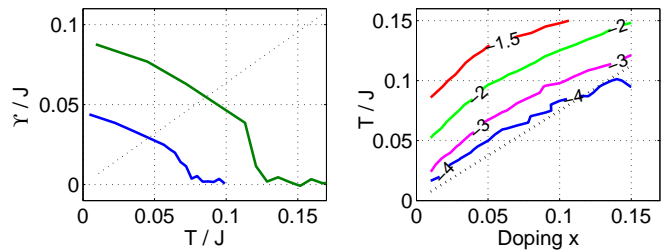}
\caption{Left: Helicity modulus $\Upsilon$ versus temperature $T$ for dopings $x=0.06$ and $x=0.12$, averaged over 300 100$\times$100 samples. The temperature where $\Upsilon$ goes to zero is an upper bound for the SC transition temperature $T_c$. The dashed line has the slope $2/\pi$. Right: Vortex density $\log_{10}n_V$ per site vs. $T$ and $x$. The dotted line is 
$T=\rho_s = 0.75xJ$.}
\label{RVplot}
\end{figure} 

In the left plot of Fig. \ref{RVplot} we show the helicity modulus $\Upsilon$ for two dopings $x$ and a 100$\times$100 system. 
$\Upsilon$  goes to zero above a doping-dependent temperature, but 
 does not exhibit the universal jump of the XY-model\cite{nelson,teitel}. 
This is clearly a finite-size effect which is exacerbated by the small vortex density and which does not occur in the XY-model where the core energy is zero. 
An estimate for the true $T_c$ is the intersection of the data with the line $2T/\pi$, that is based on the the jump criterion $\left( \Delta \Upsilon \right)_{T_c} = \pi/2$\cite{nelson}. 
With that we arrive at $T_c^e \sim 0.75 x J \sim \rho_s$. A numerical bound is
$T_c \le xJ$. 
Note that in a $d$-wave superconductor thermally excited nodal quasiparticles lead to an additional reduction of the superfluid density that is not included here. We expect however that this does not affect the nature of the SC transition.

The temperature and doping dependence of the vortex density is summarized in the right plot of Fig. \ref{RVplot}. The onset temperature for a finite vortex density is an increasing function of $x$, approximately given by $T_V \sim \rho_s$. 
In the underdoped system there is wider temperature range, starting at the BKT $T_c$ and extending up to the mean-field $T_c \sim 0.18 J$ where the vortex density continues to increase and does not saturate. Thus the vortices do not overlap and there is the possibility that SC phase coherence is still well defined {\em locally} in a range above $T_c$. For $x=0.06$ the phase correlation length $\xi_\alpha$ is  $\sim 7$ lattice spacings $a$ at $T \sim 2 T_c^e$. This is in contrast with the normal XY-model where at $2T_c$, $\xi_\alpha \le a$, and on average there is a vortex on every 5th site.
We emphasize that in our model the phase fluctuations are not the only source of disorder. This can be clearly seen in the right panel of Fig. \ref{sam}. 
The SF fluctuations are not limited to the vortex cores and lead to sizable amplitude fluctuations as well. 
Thus the proximity to the Mott state gives rise to the low energy scale for SF fluctuations and is responsible for both phase {\em and} amplitude fluctuations proliferating at comparable temperature scales.

We now address the electronic excitations.  At finite doping the bulk SF state has small Fermi pockets, and correspondingly it is natural to expect that a state which fluctuates between a superconductor with gap nodes and a SF state with small Fermi arcs will produce a finite density of states (DOS) at low energies. When we restrict the considerations to static configurations, we can calculate the quasiparticle spectrum from Hamiltonian (\ref{fermions}) for a given configuration of fluctuations on a finite system  and average over many samples.

Results are shown in the right plot of Fig. \ref{vcont}. The DOS exhibits a suppression for all temperatures below the mean-field transition at $T \sim 0.18 J$, but the gap fills in when $T$ is increased through the SC transition. The local DOS is inhomogeneous, but is not simply correlated with or confined to the positions of vortices or regions of higher SF amplitude. 
These results share some aspects with a recent work by Eckl et al.\cite{eckl} who considered disordering the $d$-wave superconductor by phase fluctuations modeled by a simple XY model. In their model the vortex cores are conventional and the core energy is zero by construction. Our generalized model accomodates both cheap vortices with SF core and energetically more expensive vortices with Fermi liquid core. 
Our calculation shows that already at the BKT $T_c$ there is a finite number of quasiparticle excitations at low energies, and it is likely that the conductivity will be dominated by these normal excitations. 
The quasiparticle contribution to the Nernst signal however is small\cite{wang} and the vortex contribution will dominate the Nernst signal at low $T$.

In conclusion, we have presented a simple model, motivated by the SU(2) approach for the high-$T_c$ cuprates, that describes the superconducting transition as an unbinding transition of vortices with staggered flux core. 
Using parameter estimates from projected wave functions and the SU(2) mean-field theory it allows us to understand the occurrence of a moderate vortex density even for underdoped systems with low $T_c$. 
The vortex density  is determined by an energy scale that is closely related to the energy difference between the SF and the $d$-wave SC state and that disappears towards zero doping, making the vortices relatively cheap. 
However there is wider temperature range above $T_c$ where the vortices are sufficiently dilute and  do not overlap such that we expect that the superconducting  phase coherence stays intact locally. This is in accordance with the interpretations of the Nernst effect measurements\cite{wang}.
Phase and SF (i.e. amplitude) fluctuations lead to a filling in of the gap in the density of states at small energies already near the  superconducting $T_c$. 
This could account for the normal-looking in-plane transport of the pseudogap phase  which onsets just above the superconducting transition. A full analysis of this issue requires a calculation of the flux-flow resistivity in the fluctuating state.   
Although our approach involves rather strong simplifications and approximations we believe that it describes a way to understand the simultaneous occurrence of normal transport behavior (e.g. in the resistivity) and strong SC fluctuations, as witnessed by the vortex Nernst effect. 

We thank D.A. Ivanov, O. Motrunich  and A. Vishwanath for useful discussions. CH acknowledges support by the German research foundation (DFG) and PAL by NSF grant DMR-0201069.

\end{document}